\documentstyle[psfig,nicV]{article}
\def\ave#1{\langle #1 \rangle}

\def\rund#1{\left( #1 \right)}

\def \ga {\mathrel{\vcenter
     {\offinterlineskip \hbox{$>$}\hbox{$\sim$}}}}
\begin{document}
%
%
\heading{%
Core Collapse Supernovae ---\\ 
Theory between Achievements and New Challenges\\   
%
}
\par\medskip\noindent
%
\author{%
H.-Thomas Janka$^{1}$
}
\address{
Max-Planck-Institut f\"ur Astrophysik, Postfach 1523,
D--85740 Garching, Germany
}
\begin{abstract}
Multi-dimensional hydrodynamic simulations of the post-bounce
evolution of collapsed stellar iron cores have demonstrated that 
convective overturn between the stalled shock and the neutrinosphere
can have an important effect on the neutrino-driven explosion 
mechanism. Whether a model yields a successful explosion or not,
however, still depends on the power of neutrino energy deposition
behind the stalled shock. The neutrino interaction with the stellar 
gas in the ``hot bubble'' also determines the 
duration of the shock stagnation phase, the explosion energy, and the
composition of the neutrino-heated supernova ejecta. More accurate
models require a more precise
calculation of the neutrino luminosities and spectra and of the 
angular distributions of the neutrinos in the heating region.
Therefore it is necessary to improve the numerical treatment of the
neutrino transport, to develop a better understanding of the 
neutrino opacities of the dense nuclear medium, and to take into
account convective processes {\it inside} the newly formed neutron star.
\end{abstract}

\section{Convective instabilities --- crucial for the explosion?}

\subsection{First hints from observations}

Supernova~1987A in the Large Magellanic Cloud, which was the nearest
visible type II supernova for more than 380 years, brought a wealth
of observational data. The neutrino measurements by the 
Kamiokande~\cite{janka.hir87}, IMB~\cite{janka.bio87}
and Baksan~\cite{janka.ale88} laboratories confirmed expectations
based on theoretical models that neutrinos play a crucial role during the
collapse of the stellar core. The photon emission from the supernova
revealed that large-scale deviations from spherical
symmetry develop during the explosion. This was suggested by the fact
that nickel clumps were seen moving at velocities much faster than 
predicted by spherically symmetric models for 
the layers where explosive nucleosynthesis of iron-group
elements takes place. Very strong mixing of hydrogen deep into the
stellar interior and of helium, metals and radioactive nuclei 
far out into the hydrogen envelope had to be invoked in order to 
reproduce the shape and the smoothness of the observed light
curve~\cite{janka.arn88,janka.shi88,janka.woo88} and to understand 
the early appearance of X-rays~\cite{janka.dot87,janka.sun87} and 
$\gamma$-rays~\cite{janka.mat88,janka.mah88,janka.coo88,janka.san88} 
from SN~1987A. Efforts to explain the large extent of the mixing
and the very high Ni velocities by hydrodynamic instabilities at 
the composition interfaces of the progenitor star 
failed~\cite{janka.arn89,janka.den90,janka.hac90,janka.her91}.
Therefore one was tempted to conclude that the anisotropies might 
originate from hydrodynamic instabilities during the very early
moments of the explosion.

%
%
%
%
%
%
%
%
%
%
%
%

\subsection{New dimensions in modeling}

This inspired multi-dimensional hydrodynamic simulations of 
neutrino-driven supernova 
explosions~\cite{janka.bur95,janka.her92,janka.her94,janka.jan95,janka.jan96,janka.lic98,janka.mez98,janka.mil93,janka.shi94} 
which indeed confirmed conjectures that the negative entropy gradient that
is built up by neutrino heating might lead to convective instabilities between
the layers of strongest energy deposition and the supernova
shock~\cite{janka.bet90}. 

The main processes of neutrino energy transfer to the stellar gas
are the charged-current reactions
$\nu_e+n \rightarrow p + e^-$ and 
$\bar\nu_e+p \rightarrow n + e^+$~\cite{janka.bet85}.
The heating rate per nucleon ($N$) is approximately given by
\begin{equation}
Q_{\nu}^+\,\approx\, 110\cdot \rund{
{L_{\nu_e,52}\langle\epsilon_{\nu_e,15}^2\rangle\over r_7^2\,\,\ave{\mu}_{\nu_e}}
\,Y_n\,+\,
{L_{\bar\nu_e,52}\langle\epsilon_{\bar\nu_e,15}^2\rangle\over r_7^2\,\,
\ave{\mu}_{\bar\nu_e}}\,Y_p
}
\quad
\left\lbrack {{\rm MeV}\over {\rm s}\cdot N}\right\rbrack \, ,
\label{janka:eq-1}
\end{equation}
where $Y_n$ and $Y_p$ are the number fractions of free neutrons and protons,
respectively, $L_{\nu,52}$ denotes the luminosity of $\nu_e$ or $\bar\nu_e$ in
$10^{52}\,{\rm erg/s}$, $r_7$ the radial position in $10^7\,{\rm cm}$,
and $\langle\epsilon_{\nu,15}^2\rangle$ the mean value of the squared neutrino
energy measured in units of $15\,{\rm MeV}$. 
$\ave{\mu}_{\nu} = \ave{\cos\theta_{\nu}}$ is the cosine of the angle 
$\theta_{\nu}$ of the
direction of neutrino propagation relative to the radial direction, 
averaged over the neutrino phase space distribution.
This quantity is very small in the opaque regime where
the neutrinos are isotropic, adopts a value of about 0.25 around the 
neutrinosphere, and approaches unity for radially streaming neutrinos
very far out.

There is general agreement about the existence and the growth of 
hydrodynamic instabilities in a layer between the shock position
and the radius of maximum neutrino heating (which
is just outside the ``gain radius'', i.e.~the radius where 
neutrino cooling turns into net heating). However, the results are less
definite concerning the question 
whether the convective overturn is strong enough to ensure the
success of the neutrino-heating mechanism in driving the explosion
of a core-collapse supernova.

%
%
\begin{figure}[t]
\centerline
{\vbox{
\psfig{figure=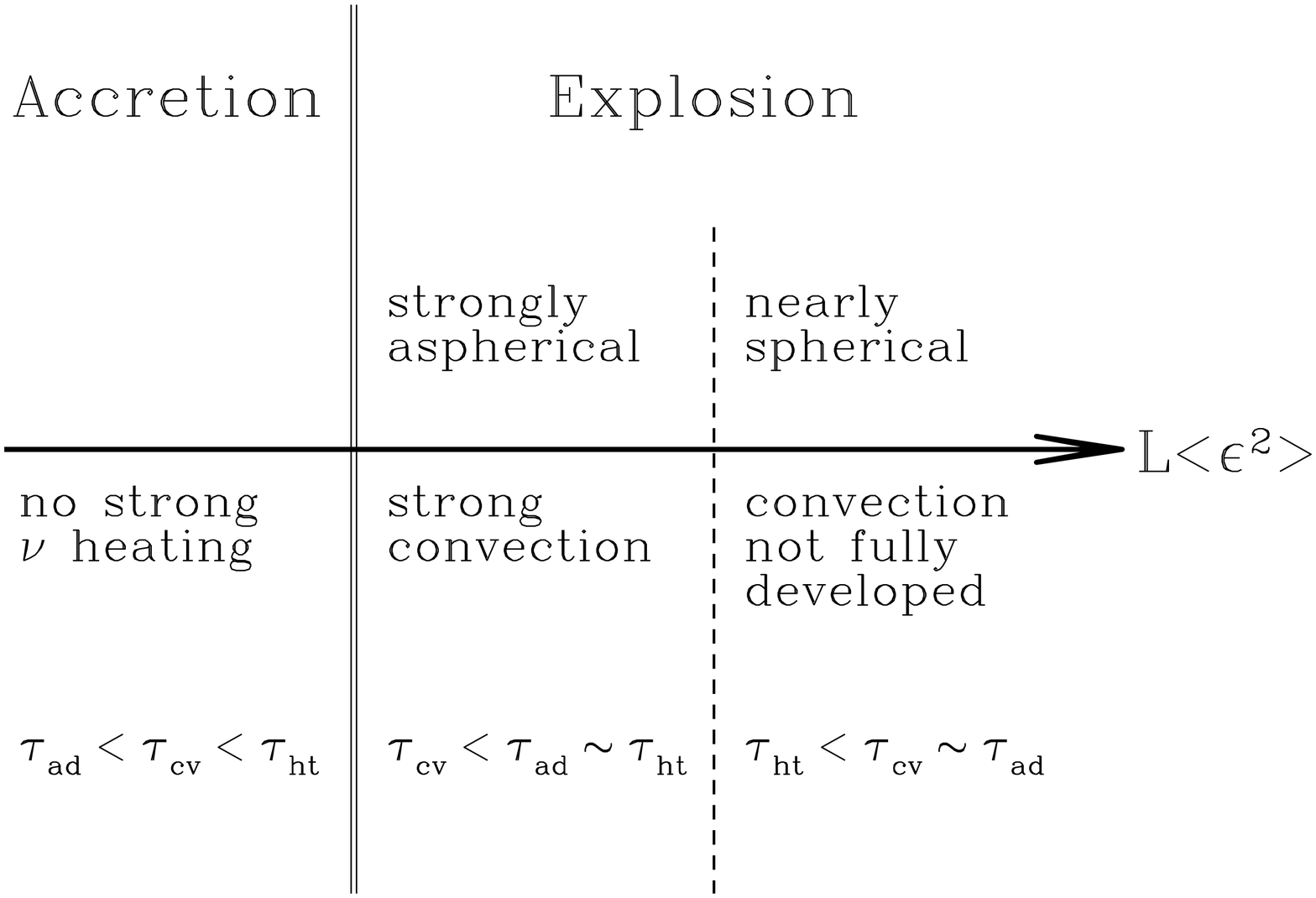,height=0.70\textwidth}}}
\caption[]{\small
Order scheme for the post-collapse dynamics in dependence of
$L_{\nu}\langle\epsilon_{\nu}^2\rangle$ which determines the strength of
the neutrino heating. The destiny of the star --
accretion or explosion -- depends on the relative size of the timescales
of neutrino heating, $\tau_{\rm ht}$, matter advection through the
gain region onto the nascent neutron star, $\tau_{\rm ad}$, and growth
of convective instabilities, $\tau_{\rm cv}$.
}
\label{janka-rev.fig3}
\end{figure}
%
%
%

\subsection{Theory between success and failure}
 
Two-dimensional simulations
by Herant et al.~\cite{janka.her94} and Burrows et al.~\cite{janka.bur95}
yielded successful explosions in cases where spherically symmetric models
fail. According to these simulations, the convective overturn in the
neutrino-heated region has the following effects on the shock propagation.
On the one hand, heated matter from the region close to
the gain radius rises outward and at the same time is replaced by cool
gas flowing down from the postshock region. Since the production reactions
of neutrinos ($e^{\pm}$ capture on nucleons and thermal processes) are
very temperature sensitive, the expansion and cooling of rising plasma
reduces the energy loss by reemission of neutrinos. Moreover, the net energy
deposition by neutrinos is enhanced as more cool material is exposed to the
large neutrino fluxes just outside the gain radius where the neutrino
heating rate peaks (the radial dilution of the fluxes roughly goes as $1/r^2$).
On the other hand, hot matter floats into the postshock region and increases
the pressure there. Thus the shock is pushed further out which leads
to a growth of the gain region and therefore also of the net energy transfer
from neutrinos to the stellar gas.

In contrast, Mezzacappa et al.~\cite{janka.mez98} and 
Lichtenstadt et al.~\cite{janka.lic98}, using multi-energy-group instead 
of a simpler grey treatment of neutrino diffusion, found that the convective 
overturn and its associated effects are not strong enough to revive the
stalled prompt supernova shock, although the outward motion of the
shock is enhanced. 

These results can be understood in view of parametric studies carried out 
by Janka \& M\"uller~\cite{janka.jan95,janka.jan96}. Varying the neutrino
luminosities from the neutrinosphere Janka \& M\"uller found that successful, 
strong explosions as well as weak ones, and even failures, can be obtained
{\it with and without} convection.
Their simulations showed that there is
a threshold luminosity above which the models explode and the explosion
energy varies sensitively with the value of the $\nu_e$ and $\bar\nu_e$ 
luminosities. 

More generally, the outcome of the simulations is determined by the
strength of the neutrino heating (minus neutrino cooling), which according
to Eq.~(\ref{janka:eq-1}), depends on $L_{\nu_e}$ and $L_{\bar\nu_e}$,
on the neutrino spectra via the mean squared neutrino energies
$\ave{\epsilon_{\nu}^2}$, and on the angular distributions of the neutrinos
through the inverse of the ``flux factors'' $\ave{\mu}_{\nu}$, 
because the neutrino number densities at radius $r$ are given by
$n_{\nu}(r) = L_{\nu}/(4\pi r^2c \ave{\epsilon_{\nu}}\ave{\mu}_{\nu})$. 
From the results of Janka \& M\"uller~\cite{janka.jan95,janka.jan96} 
one can conclude
that convective energy transport into the postshock region is crucial 
for getting an explosion only in a certain window of values of the products 
$L_{\nu}\ave{\epsilon_{\nu}^2}$ for $\nu_e$ and $\bar\nu_e$ 
(Fig.~\ref{janka-rev.fig3}). 
For higher values, the heating timescale
is so short and the energy
deposition so strong that explosions develop before convective instabilities
can grow. For lower values, neither the neutrino heating nor the convection
is efficient enough to prevent that most of the energy transferred to the stellar
gas is advected down through the gain radius and therefore lost again due to 
neutrino cooling.

%
%
%
%
%
%
%
%

\section{Routes to progress in modeling neutrino-driven explosions}

The discrepant results of different simulations and the unpleasant 
sensitivity of the explosion to the strength of the heating by $\nu_e$
and $\bar\nu_e$ absorption demands calculation of the factors in 
Eq.~(\ref{janka:eq-1}) to the highest possible accuracy. For this 
purpose, the numerical treatment of the neutrino transport in the
models needs further improvement, and the processes must be understood
in more detail which govern the emission of neutrinos from the
nascent neutron star and which thus determine the luminosities and spectra.
Current interest of supernova modelers is focused on the following 
aspects.

\subsection{Neutrino transport}

The flux factor $\ave{\mu}_{\nu}$ cannot be reliably computed by flux-limited
diffusion methods which fail to yield a good approximation of 
neutrino transport in the semi-transparent layers where neutrinos
begin to decouple from the stellar medium and where the neutrino
heating takes place. The flux factors calculated with such 
approximate methods of neutrino transport are systematically too
large~\cite{janka.mes98,janka.yam98}, 
implying that the energy deposition by
$\nu_e$ and $\bar\nu_e$ absorption according to Eq.~(\ref{janka:eq-1})
is {\it underestimated}. 

Therefore efforts are made to solve
the Boltzmann transport equation in hydrodynamic simulations of
the shock-stagnation and neutrino-heating phases. Progress is 
led by the Mezzacappa et al.~group~\cite{janka.mez93,janka.mes98} 
whose discrete ordinate (S$_N$) method provides an excellent
description of the neutrino spectra 
and a significant improvement of the angular moments of the neutrino
distribution, although the latter still deviate from 
very accurate Monte Carlo results at a level of about 10\%
when the computations are performed with an affordable number
(6--8) of angular grid points~\cite{janka.yam98}.
Since the Boltzmann solver yields values of the flux factor which
are somewhat lower than the Monte Carlo results, one expects a slight 
{\it overestimation} of the neutrino heating in this case, as can be 
verified from Eq.~(\ref{janka:eq-1}).
Mezzacappa et al.~\cite{janka.mes98}, however, claim that despite of the
remaining differences, convergence can be achieved for the net heating.

%
%
%
%

\subsection{Convection in the nascent neutron star}

The luminosities and spectra of neutrinos emitted from the 
neutrinosphere can be significantly affected by convective processes 
inside the nascent neutron
star~\cite{janka.bur87,janka.may88,janka.wil88,janka.wil93,janka.kei96,janka.kei97,janka.jan98}.
Since neutrinos can be carried to the neutrinosphere by mass motions
much faster than by the slow diffusion in the opaque interior of the 
neutron star, the two-dimensional simulations of 
Keil et al.~\cite{janka.kei96,janka.kei97,janka.jan98}
showed an amplification of the neutrino luminosity by nearly a factor
of two at about half a second after core bounce relative to spherically
symmetric models without convection.
Of course, this aids delayed explosions by neutrino heating.

Convection inside the newly formed neutron
star can be driven by gradients of the entropy and/or proton 
(electron lepton number) fraction in the nuclear
medium~\cite{janka.eps79}. The type of instability
which grows most rapidly, e.g., doubly diffusive neutron-finger
convection~\cite{janka.may88,janka.wil88,janka.wil93} or
Ledoux convection~\cite{janka.bur87} or 
quasi-Ledoux convection~\cite{janka.kei97,janka.jan98}, is
a matter of the properties of the nuclear equation of state which
determines the magnitudes and signs of the thermodynamic 
derivatives~\cite{janka.bru96}. It is also sensitive to the gradients
that develop, and thus may depend on
the details of the treatment of neutrino transport in the dense interior
of the star. 

Convection below the neutrinosphere 
seems to be disfavored during the very early post-bounce evolution by the
currently most elaborate supernova
models~\cite{janka.bru94,janka.bru95,janka.mez98}, but can develop
deeper inside the nascent neutron star on a longer timescale 
($\ga 100\,$ms after bounce) and can encompass the whole star within
seconds~\cite{janka.bur87,janka.kei96,janka.kei97,janka.jan98}. 
Recent calculations by Pons et al.~\cite{janka.pon98} for the neutrino 
cooling of hot neutron stars were done with improved neutrino opacities
of the nuclear medium which were described consistently with
the employed equation of state. Their models confirm principal
aspects of previous simulations, in particular the existence of 
convectively unstable layers in the neutron star.

\subsection{Neutrino opacities of dense matter}

Another important issue of interest are the neutrino opacities in
the dense and hot nuclear medium of the nascent neutron star. In
current supernova models, the description of neutrino-nucleon 
interactions is incomplete because the standard approximations
assume isolated and infinitely massive nucleons~\cite{janka.tub75}. 
Therefore effects like the fermion phase space blocking of the
nucleons, the reduction of the effective nucleon mass 
by momentum-dependent nuclear interactions in the dense plasma,
and nucleon thermal motions and recoil are either 
neglected completely or approximated in a more or less controled
manner~\cite{janka.bru85,janka.bur86}. These effects have
been recognized to be 
important~\cite{janka.sch90,janka.jak96,janka.pra97,janka.red97}
for reliable calculations of the neutrino luminosities and spectra,
but still await careful inclusion in supernova codes.

Many-body (spatial) correlations due to strong
interactions~\cite{janka.saw89,janka.hor91,janka.bur98,janka.red98}
and multiple-scattering effects by spin-dependent forces between 
nucleons (temporal spin-density correlations)~\cite{janka.raf95,janka.han97}
are of particular interest, because they lead to
a reduction of the neutrino opacities in the newly formed neutron star 
and are associated with additional modes of energy transfer between
neutrinos and the nuclear medium. 

A reduction of the neutrino opacities implies
larger neutrino mean free paths and thus increases the
neutrino luminosities~\cite{janka.kei95,janka.pon98,janka.bur98}. 
Despite of accelerated
neutrino diffusion, however, convection in the nascent neutron star 
turns out not to be suppressed, but to be still the fastest mode of energy 
transport~\cite{janka.jak98}. Therefore Janka et al.~\cite{janka.jak98}
found that the combined effects of reduced 
opacities and convective energy transport do not appreciably change the
convectively enhanced neutrino emission. 

Energy transfer in neutrino-nucleon collisions due to nucleon recoil and 
inelastic $\nu NN\leftrightarrow NN\nu$ scatterings are ignored in current 
supernova simulations. In addition to the nucleon bremsstrahlung 
process $\nu\bar\nu NN \leftrightarrow NN$~\cite{janka.suz93}
these effects accomplish a stronger energetic
coupling between the stellar plasma and the neutrinos and thus could 
lead to a significant 
reduction of the mean spectral energies of emitted 
neutrinos~\cite{janka.jak96,janka.han97}. In particular, it must be 
suspected that the average energies of $\bar\nu_e$ and heavy-lepton
neutrinos $\nu_{\mu},\,\bar\nu_{\mu},\,\nu_{\tau}$ and $\bar\nu_{\tau}$
on the one hand, and $\nu_e$ on the other, could be much more similar
than predicted by current supernova and neutron star cooling models.

It is not clear yet whether these discussed aspects of neutrino-nucleon 
interactions in dense nuclear matter have any implications for the 
explosion mechanism.
But due to their consequences for the neutrino luminosities and
spectra they are certainly important for the detection of neutrinos
from Galactic supernovae with future experiments like OMNIS and LAND.

\section{Summary}

Hydrodynamic simulations have shown that the shock revival phase is 
a very turbulent epoch in the supernova evolution where violent 
convective overturn takes place in the region between the neutrinosphere
and the stalled shock front. Energy transfer to the gas in this region by 
absorptions of $\nu_e$ and $\bar\nu_e$ emitted from the settling,
newly formed neutron star can deposit the energy for a powerful
supernova explosion, provided the neutrino luminosities are high
enough and/or the neutrinos have sufficiently hard spectra.

Self-consistent multi-dimensional simulations of stellar core 
collapse and of the post-bounce evolution 
carried out by different groups still yield discrepant results. While
models with a simple grey (energy-integrated) approximation of 
neutrino diffusion show successful explosions, more elaborate 
multi-energy group treatments of neutrino diffusion have so far
only produced duds.

At present it is not clear whether failures can be converted into 
successes if deficiencies of the diffusion approximation of neutrino
transport in the semi-transparent layers are removed, which lead to 
an underestimation of the neutrino energy deposition in the heating
region. Efforts are currently made to solve the Boltzmann transport
equation in combination with the equations of hydrodynamics. It is hoped 
that the more accurate treatment will yield larger neutrino luminosities
as well as higher efficiencies of neutrino energy transfer in the 
post-shock layer.

However, there are more factors of uncertainty still present
in current supernova simulations. On the one hand, neutrino-nucleon 
interactions are described by simplifying or even by ignoring a number of 
potentially very important complications, e.g., nucleon recoil and
thermal motions and nucleon phase space blocking. Other aspects like 
nucleon-nucleon correlations in the dense medium and 
multiple-scattering effects due to spin-dependent nucleon interactions
are incompletely understood.
These effects could enhance the neutrino luminosities significantly
and might also change the emitted neutrino spectra considerably.

On the other hand, recent multi-dimensional hydrodynamic simulations have 
confirmed previous conclusions from one-dimensional models that
convectively unstable layers can develop in the nascent neutron star.
The simulations show that convection can encompass the whole star
within seconds and is more efficient than neutrino diffusion in 
transporting energy to the surface. This leads to a sizable increase of 
the neutrino luminosities and to a hardening of the emitted neutrino 
spectra. If neutrino-driven explosions need several 100$\,$ms after
core bounce to develop, convective enhancement of the neutrino
luminosities could play a helpful role.

\acknowledgements{It is a pleasure to thank the organizers of NiC~V for an
enjoyable week in Volos. The author is grateful to W.~Keil, E.~M\"uller,
G.~Raffelt and S.~Yamada
for productive and pleasant collaborations. Support by the DFG
on grant ``SFB 375 f\"ur Astro-Teilchenphysik'' is acknowledged.}

\vfill

\begin{iapbib}{99}{
\bibitem{janka.ale88} Alexeyev E.N., {\it et al.}, 1988, Phys. Lett. B205, 209
\bibitem{janka.arn88} Arnett W.D., 1988, \apj 331, 337
\bibitem{janka.arn89} Arnett W.D., Fryxell B.A., M\"uller E., 1989, \apj 341, L63\\
		      Fryxell B.A., M\"uller E., Arnett W.D., 1991, \apj 367, 619
\bibitem{janka.bet90} Bethe H.A., 1990, Rev. Mod. Phys. 62, 801
\bibitem{janka.bet85} Bethe H.A., Wilson, J.R., 1985, \apj 295, 14
\bibitem{janka.bio87} Bionta R.M., {\it et al.}, 1987, Phys. Rev. Lett. 58, 1494
\bibitem{janka.bru85} Bruenn S.W., 1985, ApJS 58, 771
\bibitem{janka.bru94} Bruenn S.W., Mezzacappa A., 1994, \apj 433, L45
\bibitem{janka.bru95} Bruenn S.W., Mezzacappa A., Dineva T., 1995, Phys. Rep. 256, 69
\bibitem{janka.bru96} Bruenn S.W., Dineva T., 1996, \apj 458, L71
\bibitem{janka.bur87} Burrows A., 1987, \apj 318, L57
\bibitem{janka.bur86} Burrows A., Lattimer J.M., 1986, \apj 307, 178
\bibitem{janka.bur98} Burrows A., Sawyer R.F., 1998, Phys. Rev. C58, 554\\
		      Burrows A., Sawyer R.F., 1998, submitted to Phys. Rev. Lett. 
                      (astro-ph/9804264) 
\bibitem{janka.bur95} Burrows A, Hayes J., Fryxell B.A., 1995, \apj 450, 830
\bibitem{janka.coo88} Cook W.R., {\it et al.}, 1988, \apj 334, L87
\bibitem{janka.den90} Den M., Yoshida T., Yamada Y., 1990, Prog. Theor. Phys. 83, 723\\
		      Yamada Y., Nakamura T., Oohara K., 1990, Prog. Theor. Phys. 84, 436
\bibitem{janka.dot87} Dotani T., {\it et al.}, 1987, Nature 330, 230
\bibitem{janka.eps79} Epstein R.I., 1979, MNRAS 188, 305
\bibitem{janka.hac90} Hachisu I., Matsuda T., Nomoto K., Shigeyama T., 1990, \apj 358, L57\\
		      Hachisu I., Matsuda T., Nomoto K., Shigeyama T., 1991, \apj 368, L27
\bibitem{janka.han97} Hannestad S., Raffelt G. 1997, submitted to \apj (astro-ph/9711132)
\bibitem{janka.her91} Herant M., Benz W., 1991, \apj 345, L412\\
		      Herant M., Benz W., 1992, \apj 387, 294
\bibitem{janka.her92} Herant M., Benz W., Colgate S.A., 1992, \apj 395, 642
\bibitem{janka.her94} Herant M., Benz W., Hix W.R., Fryer C.L., Colgate S.A.,
	              1994, \apj 435, 339
\bibitem{janka.hir87} Hirata K., {\it et al.}, 1987, Phys. Rev. Lett. 58, 1490
\bibitem{janka.hor91} Horowitz C.J., Wehrberger K., 1991, Phys. Lett. B 266, 236\\
		      Horowitz C.J., Wehrberger K., 1991, Nucl. Phys. A 531, 665 
\bibitem{janka.jan98} Janka H.-Th., Keil W., 1998, eds. Labhardt L., Binggeli B.
		      and Buser R., in {\it Supernovae and Cosmology}, 
		      Astronomisches Institut der Universit\"at Basel, Basel, p. 7
		      (astro-ph/9709012)
\bibitem{janka.jan95} Janka H.-Th., M\"uller E., 1995, 1995, \apj 448, L109
\bibitem{janka.jan96} Janka H.-Th., M\"uller E., 1996, \aeta 306, 167
\bibitem{janka.jak96} Janka H.-Th., Keil W., Raffelt G., Seckel D., 1996, Phys. Rev. 
                      Lett. 76, 2621
\bibitem{janka.jak98} Janka H.-Th., Keil W., Yamada S., 1998, in preparation
\bibitem{janka.kei97} Keil W., 1997, PhD Thesis, Technical University Munich 
\bibitem{janka.kei95} Keil W., Janka H.-Th., Raffelt G., 1995, Phys. Rev. D51, 6635
\bibitem{janka.kei96} Keil W., Janka H.-Th., M\"uller E., 1996, \apj 473, L111
\bibitem{janka.lic98} Lichtenstadt I., Khokhlov A.M., Wheeler J.C., 1998,
		      submitted to \apj
\bibitem{janka.mah88} Mahoney W.A., {\it et al.}, 1988, \apj 334, L81
\bibitem{janka.mat88} Matz S.M., {\it et al.}, 1988, Nature 331, 416
\bibitem{janka.may88} Mayle R.W., Wilson J.R., 1988, \apj 334, 909
\bibitem{janka.mes98} Messer O.E.B., Mezzacappa A., Bruenn S.W. Guidry M.W., 1998,
		      \apj, Nov.~1 issue (astro-ph 9805276)
\bibitem{janka.mez93} Mezzacappa A., Bruenn S.W., 1993, \apj 405, 637\\
		      Mezzacappa A., Bruenn S.W., 1993, \apj 405, 669\\
                      Mezzacappa A., Bruenn S.W., 1993, \apj 410, 740
\bibitem{janka.mez98} Mezzacappa A., Calder A.C., Bruenn S.W., Blondin J.M.,
                      Guidry M.W., Strayer M.R., Umar A.S., 1998, \apj493, 848\\
                      Mezzacappa A., Calder A.C., Bruenn S.W., Blondin J.M.,
                      Guidry M.W., Strayer M.R., Umar A.S., 1998, \apj 495, 911
\bibitem{janka.mil93} Miller D.S., Wilson J.R., Mayle R.W., 1993, \apj 415, 278
\bibitem{janka.pon98} Pons J.A., Reddy S., Prakash M., Lattimer J.M., Miralles J.A.,
		      1998, submitted to \apj (astro-ph/9807040)
\bibitem{janka.pra97} Prakash M., {\it et al.}, 1997, Phys. Rep. 280, 1
\bibitem{janka.raf95} Raffelt G.G., Seckel D., 1995, Phys. Rev. D52, 1780\\
		      Raffelt G.G., Seckel D., Sigl G., 1996, Phys. Rev. D54, 2784
\bibitem{janka.red97} Reddy S., Prakash M., 1997, \apj 478, 689\\
		      Reddy S., Prakash M., Lattimer J.M., 1998, Phys. Rev. D58, 013009
\bibitem{janka.red98} Reddy S., Pons J., Prakash M., Lattimer J.M., 1998,
		      in Proc. Second Oak Ridge Symposium on
	  	      Nuclear and Atomic and Nuclear Astrophysics (astro-ph/9802312) 
\bibitem{janka.san88} Sandie W.G., {\it et al.}, 1988, \apj 334, L91
\bibitem{janka.saw89} Sawyer R.F., 1989, Phys. Rev. C40, 865
\bibitem{janka.sch90} Schinder P.J., 1990, ApJS 74, 249
\bibitem{janka.shi88} Shigeyama T., Nomoto K., Hashimoto M., 1988, \aeta 196, 141
\bibitem{janka.shi94} Shimizu T., Yamada S., Sato K., 1994, \apj 432, L119
\bibitem{janka.sun87} Sunyaev R.A., {\it et al.}, 1987, Nature 330, 227
\bibitem{janka.suz93} Suzuki H., 1993, eds. Suzuki Y. and Nakamura K.,
                      in {Frontiers of Neutrino Astrophysics}, 
		      Universal Academy Press, Tokyo, p. 219
\bibitem{janka.tub75} Tubbs D.L., Schramm D.N., 1975, \apj 201, 467
\bibitem{janka.wil88} Wilson J.R., Mayle R.W., 1988, Phys. Rep. 163, 63
\bibitem{janka.wil93} Wilson J.R., Mayle R.W., 1993, Phys. Rep. 227, 97
\bibitem{janka.woo88} Woosley S.E., 1988, \apj 330, 218
\bibitem{janka.yam98} Yamada S., Janka H.-Th., Suzuki H., 1998, subm.~to \aeta
		      (astro-ph/9809009)
}
\end{iapbib}
\end{document}